# Slowing down of water dynamics in disaccharide aqueous solutions

by

A. Lerbret[a] †, F. Affouard[a], P. Bordat[b],

A. Hédoux[a], Y. Guinet[a], M. Descamps[a]

[a] Unité Matériaux Et Transformations, UMR CNRS 8207

Université Lille 1, 59655 Villeneuve d'Ascq Cedex, France.

[b] Institut Pluridisciplinaire de Recherche sur l'Environnement et les Matériaux,

UMR CNRS 5254, Université de Pau et des Pays de l'Adour,

2 Avenue Pierre Angot, 64053 Pau Cedex 9, France.

† author to whom correspondence should be addressed, adrien.lerbret@univ-lille1.fr.

**Abstract**

The dynamics of water in aqueous solutions of three homologous disaccharides, namely trehalose, maltose and sucrose, has been analyzed by means of molecular dynamics simulations in the 0-66 wt % concentration range. The low-frequency vibrational densities of states (VDOS) of water were compared with the susceptibilities $\chi''$ of 0-40 wt % solutions of trehalose in $D_2O$ obtained from complementary Raman scattering experiments. Both reveal that sugars significantly stiffen the local environments experienced by water. Accordingly, its translational diffusion coefficient decreases when the sugar concentration increases, as a result of an increase of water-water hydrogen bonds lifetimes and of the corresponding activation energies. This induced slowing down of water dynamics, ascribed to the numerous hydrogen bonds that sugars form with water, is strongly amplified at concentrations above 40 wt % by the percolation of the hydrogen bond network of sugars, and may partially explain their well-known stabilizing effect on proteins in aqueous solutions.

## 1 – Introduction

Disaccharides ($C_{12}H_{22}O_{11}$) such as trehalose and sucrose are well-known for their high efficiencies in preserving biological molecules against thermal and dehydration stresses [1,2,3,4] and are accordingly widespreadly used in various industrial processes for the long-term conservation of therapeutic proteins, food and cosmetics [5,6,7]. However, the detailed molecular mechanisms underlying their bioprotecting abilities, and in particular the superior efficiency of trehalose, are still poorly understood [3]. Many hypotheses [4,8,9,10,11,12] (vitrification, replacement of hydration water molecules, etc.) have been proposed in the literature, but they generally cover only narrow temperature and/or hydration ranges that prevent a comprehensive description of the whole biopreservation mechanism.

In rather dilute aqueous solutions, various experimental [12] and simulation [13,14] results suggest that sugars are preferentially excluded from the surface of globular proteins. Therefore, their bioprotective effect is thought to partially stem from the significant slowing down they induce on the dynamics of protein hydration water molecules [15,16,17]. Indeed, sugars form numerous hydrogen bonds (HBs) with water [18,19], whose dynamics is strongly slow down in their hydration shell [15,20,21,22]. Recently, disaccharides were shown to decrease the flexibility of lysozyme [17,23], thereby reducing its conformational entropy. This result fully agrees with the stabilization of the tertiary structure of lysozyme at high temperatures observed in Raman scattering experiments [23] and would explain why the secondary structure of lysozyme unfolds at higher temperatures in presence of sugars [23,24]. Furthermore, trehalose exhibited superior protecting capabilities [23,24] that would partially stem from its higher ability to slow down water dynamics [25], mainly because of its larger hydration number [18,26,27].

In this paper, the dynamics of water in aqueous solutions of three homologous disaccharides, namely trehalose, maltose and sucrose, is investigated by means of molecular dynamics (MD) simulations in a broad range of concentrations (0-66 wt %). The low-frequency vibrational densities of states (VDOS) of water were calculated and compared with the susceptibilities χ" of 0-40 wt % solutions of trehalose in $D_2O$ obtained from complementary Raman scattering experiments. Moreover, the translational diffusion coefficient of water and the activation energies of water-water hydrogen bonds were determined to further examine the influence of the percolation of the hydrogen bond network (HBN) of sugars on the dynamical slowing down of water.

## 2 – Simulation and Experimental Details

### 2.1 Molecular Dynamics Simulations

The simulations of disaccharide aqueous solutions have been thoroughly described in ref. [27,28] and are summarized in the following. MD simulations of 512 water molecules in presence of 0, 1, 5, 13, 26 or 52 sugar molecules (either trehalose, maltose or sucrose) leading to weight concentrations of 0, 4, 16, 33, 49 or 66 %, respectively, have been performed at temperatures ranging from 273 up to 373 K in steps of 20 K using the DL_POLY software [29]. Water and sugar molecules were represented with the rigid SPC/E model [30] and with the fully flexible all-atom carbohydrate force field developped by Ha et al. [31], respectively. Electrostatic interactions were handled by the reaction-field method [32] (with $\varepsilon_{RF}=72$). The cutoff radius for non-bonded interactions was set to 10 Å and cubic periodic boundary conditions were applied. Simulations were performed in the isobaric-isothermal NPT ensemble using weak couplings to pressure and heat baths [33], with reference pressure set to 1.0 bar. Simulation times ranged from 0.2 up to 2.0 ns depending on the temperature and on the sugar concentration considered. Time steps of 0.5 and 2 fs were used to integrate the equations of motions for the binary solutions and for pure water, respectively.

### 2.2 Raman Scattering Experiments

High-purity anhydrous trehalose was supplied from Fluka and Sigma. Measurements were performed on trehalose in deuterated water at different weight fractions of sugar (0, 10, 20, 30 and 40 wt %) and T=295 ± 0.1 K. The mixtures were loaded in hermetically closed Hellma quartz Suprasil cells. The 514.5 nm line of a mixed argon-krypton laser was used for Raman excitation. The back-scattering Raman spectra were recorded in the 10-300 cm$^{-1}$ spectral window using a Dilor-XY spectrometer equipped with a liquid nitrogen cooled charge-coupled-device detector. The scattered low-frequency intensity was transformed into Raman susceptibility χ" using a procedure detailed in previous studies [23]. χ"(ν) is related to the VDOS g(ν) by the relation $\chi''(\nu) = C(\nu) \cdot \nu^{-1} \cdot g(\nu)$, where C(ν) is the light-vibration coupling coefficient. Assuming a linear frequency dependence of C(ν) [34], χ"(ν) is roughly representative of the VDOS, provided that the quasielastic contribution arising from anharmonic motions has been accurately subtracted from the low-frequency spectrum. All spectra were finally normalized between 10 and 120 cm$^{-1}$.

## 3 – Results

### 3.1 Low-frequency vibrational dynamics

The influence of sugars on the dynamics of water was first probed with the low-frequency VDOS of water, shown for the different trehalose solutions at 293 K in Fig. 1. For comparison, the low-frequency Raman susceptibilities χ" of trehalose/D2O solutions in the 0-40 % concentration range are also displayed. A qualitative agreement is obtained between MD and Raman spectra, in which two main bands appear near 50-60 and 170-200 cm$^{-1}$, respectively, though with different positions, amplitudes and widths. These discrepancies may arise from (i) the limited accuracy of force fields used in simulations, (ii) the significant contribution of sugars to the Raman spectra, as observed for frequencies above 250 cm$^{-1}$ for instance, (iii) the probably nonlinear frequency dependence of the light-vibration coupling coefficient C(ν) [17] and (iv) the nontrivial subtraction of quasielastic and fluorescence contributions. The first band, close to 50-60 cm$^{-1}$ in neat water, has been ascribed to the intermolecular vibrations that water molecules experience within the *cage* formed by their neighbors, and has also been observed both in hydrogen-bonded and nonassociated liquids [35,36]. It broadens with the addition of sugars, which may reflect the increased heterogeneity of the local environments sampled by water molecules in mixed solutions. The second band near 170-200 cm$^{-1}$ has been previously assigned to collective intermolecular O-H$^{..}$O stretching vibrations [37]. The decrease of its amplitude when the sugar concentration increases reflects the *destructuring effect* of sugars on the HBN of water [9,19,28,38], which essentially stem from the numerous HBs they form with water [27]. Furthermore, the concomitant increase in the frequency position of the two bands indicates that the local environments experienced by water molecules are stiffer in presence of sugars, in line with the sterical restriction imposed by their rather rigid skeletons [39]. In particular, the sugar-induced blue shift of the second band implies that water-water HBs are *strengthened*, as will be shown more explicitely in section 3.3.

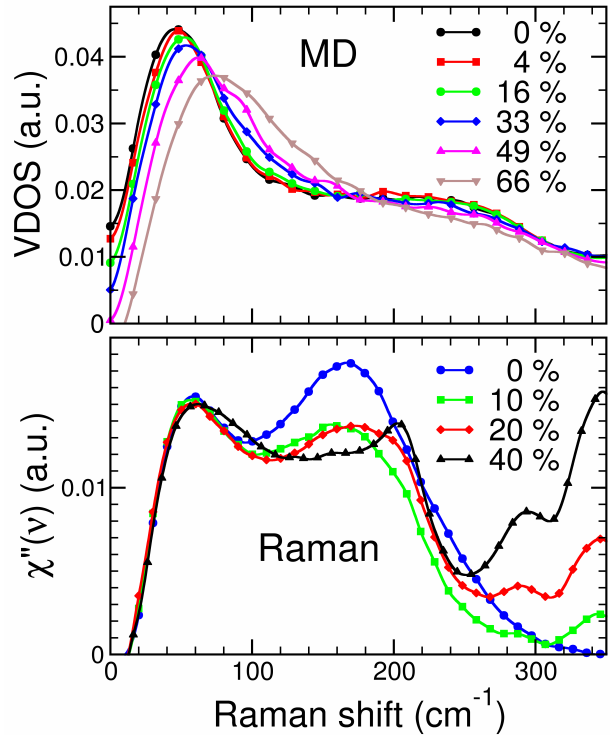

**Fig. 1 Top: Low-frequency vibrational density of states (VDOS) of water in the different trehalose/water solutions at 293 K. Bottom: Raman susceptibility $\chi''(\nu)$ of trehalose/D$_2$O solutions in the 0-40 wt % sugar concentration range at 295 K, normalized in the 10-120 cm$^{-1}$ range. MD and Raman spectra have been smoothed with the Savitzky-Golay algorithm [40] to simplify the comparison of results.**

A fit of the computed VDOS of water in the different sugar solutions has been arbitrarily performed with a log-normal and a Gaussian functions, and the dependence on the sugar concentration of the frequency positions of the two bands is shown in Fig. 2. For comparison, a similar fitting procedure was performed on the Raman susceptibilities of the trehalose solutions in D$_2$O. It must be pointed out, however, that a second gaussian function was necessary to represent satisfactorily the second band for trehalose concentrations greater or equal to 20 %, given the emergence of an additional contribution near 210 cm$^{-1}$ that may reveal the formation of particularly strong water-trehalose HBs. As seen in Fig. 1, the frequency positions $\nu_{caging}$ and $\nu_{O-H\cdots O}$ obtained from MD and Raman spectra differ significantly with each others. Nevertheless, their dependence on trehalose concentration appears in satisfying agreement, and an overall stiffening of local water environments is clearly observed. Besides, the high-frequency shifts of the two positions are found similar in the different sugar solutions. Given the large uncertainties on peak positions, an accurate comparison between sugars is not possible, even though sucrose seems to stiffen to a lower extent the environment of water molecules in the 66 wt % solution, in line with previous observations [17]. Moreover, a steep frequency increase in the position of the two bands appears for concentrations above about 40 % and is ascribed to the percolation of the HBN of sugars, which was found to induce significant changes in the structural and dynamical properties of disaccharide/water solutions [27,28].

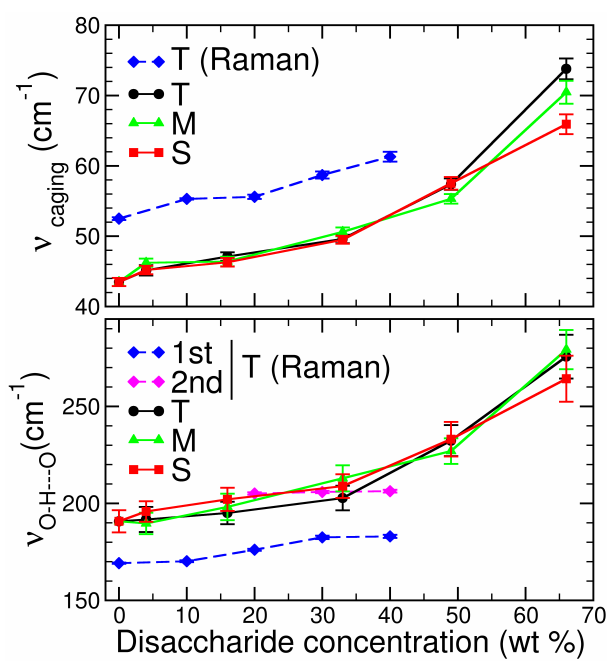

**Fig. 2** Frequency positions of the two bands of the low-frequency VDOS of water in the different sugar/water solutions at 293 K obtained by fitting in the 0-300 cm$^{-1}$ range the calculated spectra with a log-normal and a gaussian functions, respectively. These arbitrary functions were chosen since they provided the best fits. For comparison, the frequency positions deduced from similar fits of the low-frequency Raman susceptibilities χ" of trehalose solutions in heavy water are also shown. Two gaussian functions were needed to represent the second band of the experimental spectra for concentrations greater or equal to 20 %.

## 3.2 Water translational diffusion coefficients

The translational diffusion coefficients of water $D_w$ were computed from the long-time limit slope of the water mean square displacement (MSD) according to the Einstein relation $D_w = \lim_{t \to \infty} \langle |r(t) - r(0)|^2 \rangle / 6t$, where *r(t)* and *r(0)* are the position vectors of the center of mass of molecules at times *t* and 0 respectively, and the brackets means averaging over every time origin and water molecule. The calculated diffusion coefficients $D_w$ for the different disaccharide solutions at 293 K are displayed in Fig. 3. The computed $D_w$ compare well with the diffusion coefficients interpolated from the experimental data of Mills [41] for pure water and of Rampp *et al*. [39] for trehalose solutions, thereby suggesting that the influence of sugars on water dynamics is rather well reproduced in the present simulations. Water diffusion is slowed down by more than one order of magnitude when sugar concentration increases from 4 to 66 wt %. Interestingly, the concentration dependence of $D_w$ mimicks that of the probability of water HB formation $p_{HB}$ [27] or that of the mean size of water clusters $<n_W>$ [28]. This shows that the slowing down of water dynamics stems from the increasing number of water-sugar HBs as sugar concentration increases. Indeed, carbohydrates induce a significant decrease of the translational and rotational diffusion coefficient of water in their hydration shell [20], as could be expected from their significantly smaller diffusion coefficient with comparison to water [39]. The steep decrease of $D_w$ when the sugar concentration increases from 49 to 66 wt % can be ascribed to the percolation of the HBN of sugars in this concentration range [27]. It is also worth noticing that the differences between the influence of sugars on water dynamics are essentially observed at 66 wt %, at which differences between the effects of sugars on the HBN of water are significant [27,28]. At low sugar concentrations, the total number of water molecules involved in HBs with sugars is too low for the differences between the respective hydration numbers of sugars to be observed on $D_w$. These differences emerge on $D_w$ only at 66 wt %, at which the numbers of water-water and water-sugars HBs are comparable. Then, $D_w$ is found the lowest in the trehalose solution, as could be expected from its larger hydration number [18,26,27].

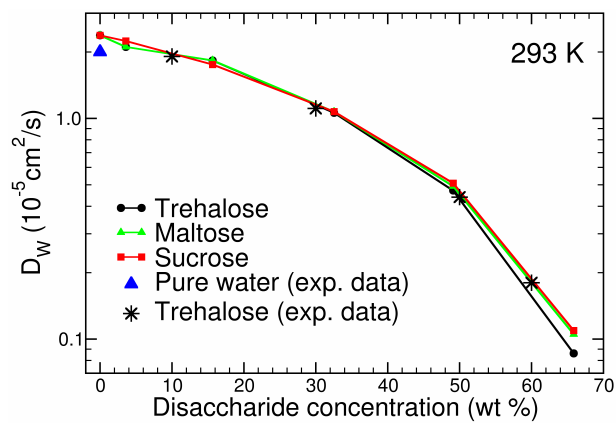

**Fig. 3** Water translational diffusion coefficient $D_w$ as a function of the disaccharide concentration in the different sugar/water solutions at 293 K. For comparison, the diffusion coefficients of water obtained from a linear interpolation of experimental data at 288 and 298 K from Mills [41] for pure water and from Rampp et al. [39] in 10, 30, 50 and 60 wt % trehalose solutions are also reported as a triangle and stars symbols, respectively.

**3.3 Water hydrogen bonds dynamics**

The dynamics of water molecules primarily depends on the dynamics of the hydrogen bonds they form with their neighbors. Here, the structural relaxation times of water-water HBs $\tau_{HB}$ have been estimated from the long-time decay of the intermittent correlation function $C(t)=\langle h_i(t).h_i(0)\rangle$ [42] where $h_i(t)$ is unity if a given HB *i* formed at time 0 remains intact at time *t* - even if it has broken in between - and is zero otherwise (two water molecules were considered to be hydrogen bonded if the O-O distance is less than 3.4 Å and if the O-H...O angle is larger than 120° [43]). Following a fast decrease of C(t) at short times (< 0.2 ps) ascribed to water libration, the long-time decay of C(t) reflects the relative diffusion of two water molecules initially H bonded. The average lifetime of water-water HBs $\tau_{HB}$ is thus directly linked to the diffusion coefficient of water molecules. The $\tau_{HB}$ strongly increase with the concentration of sugars (see for example the inset of Fig. 4), in good agreement with the water diffusion coefficients $D_w$ shown in Fig. 3. This effect stems from the greater stability of carbohydrate-water HBs compared to water-water HBs [20] and is well in line with the significant high frequency shift of the second band of the water VDOS induced by sugars (Fig. 1 and Fig. 2), interpreted as a *strengthening* of water HBs. To further characterize the strong slowing down of water HBs dynamics induced by sugars, the corresponding activation energies $E_a$ have been determined, assuming an Arrhenius behavior of $\tau_{HB}$ in the range of temperatures considered (273-373 K). At low concentrations (0-33 wt %), the $E_a$ energies range between 3.4 and 3.8 kcal.mol$^{-1}$, in good agreement with experimental data for neat water [44,45] and for moderately concentrated carbohydrate solutions [46,47,48]. At variance, the $E_a$ energies significantly increase for disaccharide concentrations above about 40 wt %, following the percolation of the sugars HBN [27]. In sharp contrast with the present results, Di Fonzo *et al.* [49] found by means of Brillouin scattering experiments an activation energy seemingly constant in trehalose/water mixtures over a broad concentration range (0-74 wt %). However, the activation energies for water diffusion in maltose glasses (concentrations above 90 wt %) were found to be about 15-17 kcal.mol$^{-1}$ [50,51]. Therefore, the concentration dependence seen in Fig. 4 seems realistic. Consistent with the calculated $D_W$ (Fig. 3), the water-water HBs activation energy is found the highest in the trehalose solution at 66 wt %, even though the related large error bars prevent any statistically meaningful comparison between the three sugars.

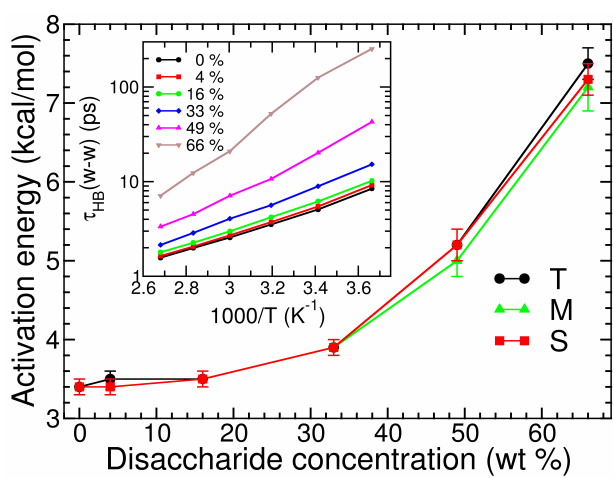

**Fig. 4 Activation energy of water-water hydrogen bonds as a function of disaccharide concentration in the different sugar/water solutions. The inset shows the temperature dependence of water-water hydrogen bonds lifetimes $\tau_{HB}$(w-w) in the different trehalose solutions.**

## 4 – Discussion

In agreement with other studies [15,20,21,22,25,39,46], the results reported here all reveal that disaccharides induce a significant slowing down of water dynamics, which depends primarily on the total number of HBs they form with water [15,22,25]. Indeed, several recent investigations clearly showed a strong retardation of the dynamics of water molecules hydrating sugars [15,20,21,22], up to distances of 5.5-6.5 Å for disaccharides (trehalose, sucrose, lactose), that is, beyond their first static solvation layer. In particular, water-water HB lifetimes were found to increase by as much as 40 % compared to the bulk in the vicinity of trehalose and lactose [15,21], because sugar-water HBs are more stable than water-water HBs [20]. In line with those results, Paolantoni *et al.* identified two relaxation processes for water in trehalose aqueous solutions, one assigned to bulk water, and the other to sugar hydration water, characterized by relaxation times 5-6 times longer [22]. The present results further underline the influence of the percolation of the HBN of sugars [27] on the average slowing down of water dynamics, which steeply increases for concentrations above about 40 wt %. This concentration corresponds approximately to the concentration threshold for which the mean cluster size of H-bonded water molecules steeply decreases [28] and for which slight differences between the respective influence of sugars on water start to appear [27,28]. At a concentration of 66 wt %, trehalose seems to slow down more efficiently the diffusion of water molecules (Fig. 3) and to increase to a larger extent the activation energies of water-water HBs (Fig. 4). This could be ascribed to the ability of trehalose to form large clusters with itself while still interacting with more water molecules than maltose and sucrose do [27]. This peculiar balance between sugar-sugar and sugar-water interactions directly stems from the topology of trehalose, whose nearly symmetric conformation [52] prevents extensive internal H-bonding between the two glucose rings. At variance, sucrose, and to a lower extent maltose, form more frequently internal HBs that reduce their abilities for intermolecular H-bonding [27]. Besides, the slowing down of water dynamics induced by sugars could explain in part their stabilizing effect on proteins at high temperatures [23,24]. Given that they are preferentially excluded from the surface of proteins in aqueous solutions [12,13,14], sugars may strongly slow down protein dynamics by forming HBs with the protein hydration water molecules [14] and then hinder the softening of protein vibrational modes that occurs upon denaturation [17]. In other words, sugars would reduce the conformational entropy of proteins by making them less flexible [23], particularly at high temperatures, and thus increase their thermal denaturation temperature $T_m$ [23,24]. This suggested retardation of the dynamics of protein hydration water [15,16] is, however, probably not the only bioprotective effect that sugars induce on proteins. For example, they may also sterically hinder protein aggregation, a major cause of instability in solutions [53]. Consequently, this hypothesis explains partially only how proteins may be stabilized by sugars in solution.

**5 – Conclusion**

The low-frequency VDOS of water and Raman susceptibilities of trehalose/D2O solutions indicate that the local environments sampled by water molecules are strongly stiffened in presence of the three studied disaccharides. As a consequence, water diffusion significantly slows down when the sugar concentration increases, following an increase of water-water hydrogen bonds lifetimes and of the associated activation energies. This induced retardation of water dynamics stems from the numerous HBs that sugars form with water and is strongly amplified at concentrations above 40 wt % by the percolation of the HBN of sugars [27]. Since sugars are known to be preferentially excluded from the surface of globular proteins [12,13,14], their stabilizing effect at high temperatures may originate from the dynamical slowing down of the protein hydration water [15,16].

**Acknowledgements**

The authors wish to acknowledge the use of the facilities of the IDRIS (Orsay, France) and the CRI (Villeneuve d'Asq, France), where calculations were carried out. This work was supported by the INTERREG III (FEDER) program (Nord-Pas de Calais/Kent) and by the ANR (Agence Nationale de la Recherche) through the BIOSTAB project ("Physique-Chimie du Vivant" program). A.L. thanks the Nord-Pas de Calais region for a postdoctoral fellowship.